\documentclass[conference]{IEEEtran}
\IEEEoverridecommandlockouts
\usepackage{cite}
\usepackage{amsmath,amssymb,amsfonts}
\usepackage{algorithmic}
\usepackage{graphicx}
\usepackage{textcomp}
\usepackage{xcolor}
\usepackage{enumitem}
\usepackage{subcaption}
\usepackage{orcidlink}
\usepackage{balance}
\usepackage{xurl} 
\usepackage{hyperref}  
\usepackage{breakurl} 
\begin{document}

\title{Validating Terrain Models in Digital Twins\\for Trustworthy sUAS Operations}

\author{\IEEEauthorblockN{Arturo Miguel Russell Bernal
\orcidlink{0009-0009-2902-5766}}
\IEEEauthorblockA{\textit{Computer Science and Eng.} \\
\textit{Univ. of Notre Dame}\\
South Bend, IN \\
arussel8@nd.edu}
\and
\IEEEauthorblockN{Maureen Petterson \orcidlink{0009-0008-8801-0751}}
\IEEEauthorblockA{\textit{Computer Science and Eng.} \\
\textit{Univ. of Notre Dame}\\
South Bend, IN \\
mpetters@nd.edu}
\and
\IEEEauthorblockN{Pedro Antonio Alarcon Granadeno  
\orcidlink{0009-0006-7829-7088}}
\IEEEauthorblockA{\textit{Computer Science and Eng.} \\
\textit{Univ. of Notre Dame}\\
South Bend, IN \\
palarcon@nd.edu}
\and
\IEEEauthorblockN{Michael Murphy 
\orcidlink{0009-0007-4303-412X}}
\IEEEauthorblockA{\textit{Computer Science and Eng.} \\
\textit{Univ. of Notre Dame}\\
South Bend, IN \\
murphym18@gmail.com}
\and
\IEEEauthorblockN{James Mason  \orcidlink{0009-0005-9542-1856}} 
\IEEEauthorblockA{\textit{Computer Science and Eng.} \\
\textit{Univ. of Notre Dame}\\
South Bend, IN \\
jmason23@nd.edu}
\and
\IEEEauthorblockN{Jane Cleland-Huang  
\orcidlink{0000-0001-9436-5606}}
\IEEEauthorblockA{\textit{Computer Science and Eng.} \\
\textit{Univ. of Notre Dame}\\
South Bend, IN \\
janehuang@nd.edu}
}

\maketitle

\begin{abstract}
With the increasing deployment of small Unmanned Aircraft Systems (sUAS) in unfamiliar and complex environments, Environmental Digital Twins (EDT) that comprise weather, airspace, and terrain data are critical for safe flight planning and for maintaining appropriate altitudes during search and surveillance operations. With the expansion of sUAS capabilities through edge and cloud computing, accurate EDT are also vital for advanced sUAS capabilities, like geolocation.  However, real-world sUAS deployment introduces significant sources of uncertainty, necessitating a robust validation process for EDT components.  This paper focuses on the validation of terrain models, one of the key components of an EDT, for real-world sUAS tasks.  These models are constructed by fusing U.S. Geological Survey (USGS) datasets and satellite imagery, incorporating high-resolution environmental data to support mission tasks. Validating both the terrain models and their operational use by sUAS under real-world conditions presents significant challenges, including limited data granularity, terrain discontinuities, GPS and sensor inaccuracies, visual detection uncertainties, as well as onboard resources and timing constraints. We propose a 3-Dimensions validation process grounded in software engineering principles, following a workflow across granularity of tests, simulation to real world, and the analysis of simple to edge conditions. We demonstrate our approach using a multi-sUAS platform equipped with a Terrain-Aware Digital Shadow.
\end{abstract}


\section{Introduction}
As swarms of small Unmanned Aircraft Systems (sUAS) are increasingly deployed in complex, unstructured environments such as disaster zones, wilderness areas, and wildfire regions, the need for accurate environmental models becomes critical. Effective sUAS mission planning requires awareness not only of dynamic airspace and weather conditions but also of the underlying terrain. In such settings, terrain is often the dominant factor influencing flight safety, sensor placement, line-of-sight communications, and search effectiveness.

This paper focuses specifically on the role of terrain models that enable mission-level decision-making and flight planning for sUAS operations. However, terrain inaccuracies or blind spots, such as missing elevation data, undetected peaks, or mismatched georeferencing, can result in ineffective or even hazardous behavior by autonomous vehicles.
To minimize these issues, we construct and maintain a terrain model by fusing multiple sources of environmental data, including public USGS datasets \cite{usgs_annual_nlcd_2025,usgs_nhdplushr}, and satellite imagery \cite{granadeno2025landcoverageawarepathplanningmultiuav}. 
The resulting Terrain-Aware Digital Shadow (TDS) provides a unidirectional mapping from the physical world to a digital representation \cite{Tekinerdogan23}, supporting real-time queries from autonomous systems, enabling proactive monitoring of terrain conditions, and serving as a runtime referential model for planning and situational awareness.
The TDS is also part of a broader ecosystem that functions as a true Digital Twin \cite{Tekinerdogan23}, providing a fully integrated, bidirectionally synchronized environment in which the physical sUAS and their virtual counterparts continuously exchange data, enabling real-time monitoring, control, and adaptive feedback between the simulated and real-world systems.
This bi-directional integration allows operators to simulate missions, dispatch commands to physical sUAS, and receive real-time telemetry updates, all from within the digital twin interface. As a result, while the TDS provides a reflective environmental model, the overall architecture enables a fully interactive twin that unifies modeling, planning, and control.

The primary focus of this paper is to examine the real-world challenges involved in constructing, deploying, and validating a TDS in support of sUAS operations. Our core objective is to rigorously evaluate how these models perform under realistic operating conditions, where uncertainty is pervasive. We test their effectiveness as part of an operational sUAS system, focusing not only on the accuracy of the models themselves, but also on how they shape and influence autonomous control, mission planning, and human oversight.

In particular, we consider three key sources of uncertainty: (1) errors in sUAS geolocation and altitude readings due to GPS limitations, especially in rugged or obstructed environments, (2) inaccuracies in the attitude estimation of the sUAS and its gimbal, which affect the alignment and projection of sensed data onto the terrain model, and (3) limitations in the underlying terrain data, caused by the coarse granularity of USGS datasets, potentially leading to the absence of critical features like narrow crevices, abrupt ridges, or localized elevation changes. Our primary goal is to plan smooth, altitude-aware missions that minimize abrupt maneuvers, preserve stable sensor views for search, provide accurate geolocation of objects of interest, and reduce crash risk through safe separation from terrain. 
To support this goal, we propose a systematic testing framework for identifying and analyzing key sources of uncertainty within the sUAS control and modeling pipeline, and for validating the resulting system behavior through structured field testing. 

Our primary contributions are therefore as follows:

\begin{enumerate}
    \item We present a Terrain-Aware Digital Shadow (TDS) designed for sUAS operations, with emphasis on runtime integration and operational use, thereby bridging the validation gap between simulation-based modeling and live field deployment.
    \item We identify and analyze challenges and key sources of model fragility, including terrain resolution, sensor alignment, and environmental variability.
    \item We propose a 3-Dimensional testing framework that validates the TDS across unit-to-system-level, simulation-to-reality, and simple-to-complex mission scenarios.
    \item We share lessons learned through the application of our framework in a use case scenario.
\end{enumerate}

The remainder of this paper is structured as follows. Section \ref{sec:modeling} briefly describes our methodology for constructing terrain models using a merged pipeline that integrates public USGS data and satellite imagery. Section \ref{sec:challenges} describes operational challenges of using the TDS. Sections \ref{sec:framework} and \ref{sec:applied} present our testing framework and describe its use for validating the TDS in the sUAS domain. Finally, Sections \ref{sec:threats} to \ref{sec:conclusions} discuss threats to validity, related work,  and draw conclusions.

\section{Constructing the TDS}
\label{sec:modeling}
A TDS is constructed for a specific flight region by retrieving and merging several USGS (United States Geological Survey) datasets, acquiring satellite imagery, segmenting it into terrain features using a trained computer vision (CV) model, and then integrating both data sources into a unified model. This two-pronged approach combines the elevation data from USGS, as a source of critical geometric and altitude information, with the up-to-date semantic detail extracted from satellite imagery.

\subsection{USGS-derived Data Models}
We utilized the following publicly available USGS datasets to build the TDS model:

\begin{itemize}
\setlength{\itemsep}{0.2em}
\item Digital Elevation Models (DEMs), which are available as raster datasets in resolutions ranging from 1 m to 10 m, depending on geographic location and data availability. Our implementation uses a Digital Terrain Map (DTM), which provides a bare-earth topographic surface, including water sources, and therefore does not include impermanent features such as vegetation canopy and structural features \cite{thatcher2020usgs}.
\item Hydrographic features such as lakes, streams, and reservoirs are available in the high-resolution National Hydrography Dataset (NHDPlusHR) with resolutions $\leq 10\ m$ \cite{usgs_nhdplushr}. 
\item Man-made features, including roads and trails, are available as vector datasets in the Transportation Dataset \cite{usgs_transportation_2023}. 
\item Categorical land cover classifications are provided by the National Land Cover Database (NLCD) at a resolution of 30 m. The 20 classifications include forest ecosystems, shrublands, grasslands, aquatic systems, wetlands, and developed surfaces \cite{usgs_annual_nlcd_2025}.
\end{itemize}

\subsection{Model Construction}
In the first step of constructing the terrain model, we define a region of interest using a grid of coordinates in the WGS84 coordinate reference system. USGS datasets for this region are then retrieved using the open-source HyRiver Python package, which provides a unified, high-level API for accessing The National Map’s 3DEP elevation service, the NHDPlusHR hydrography dataset, and NLCD land cover data \cite{chegini2021hyriver}. Transportation data is acquired separately from the AWS Staged Products directory in Shapefile format, and downloaded on a state-by-state basis.

These datasets are spatially indexed using an STRTree, a query-only R-Tree structure that organizes geometries using the sort-tile-recursive algorithm \cite{leutenegger1997str}. Continuous features such as elevation and land cover are stored in one STRTree, while discrete features—including water bodies, trails, and roads—are stored in a separate structure. For each pixel in the region of interest, the corresponding (latitude, longitude) is used to query the STRTrees and extract relevant geospatial attributes. Discrete feature extraction is performed using geometric intersection algorithms, implemented in Shapely and derived from GEOS and the Java JTS Topology Suite \cite{geos2024, jts_martindavis, jts_locationtech}. These intersections are determined using intersection matrices that evaluate spatial relationships between geometries.

For continuous data, the nearest neighbor node in the STRTree is identified using a Cartesian 2D distance metric. The selected node returns the elevation and land cover classification corresponding to the closest available data point. Although the current approach is based on centroid-based queries, future enhancements may incorporate radius-based searches to improve robustness under spatial uncertainty \cite{leutenegger1997str}. The resulting terrain model is stored in a nested dictionary structure, where each 2D cell contains its corner coordinates, centroid, elevation, and land cover classification. Discrete features are appended as attributes associated with each cell.

The resolution of the terrain model is adjustable through the grid cell size, though it remains ultimately limited by the resolution and completeness of the source datasets. For instance, earlier efforts using AWS-staged land cover data encountered missing tiles, prompting a transition to HyRiver to improve data coverage and integration reliability. The final terrain model supports both direct geospatial queries and the generation of composite visualizations using colormaps to represent terrain features.

In the second step, a high-resolution satellite image of the selected region is retreived via an external imagery API, (e.g., Mapbox API \cite{mapbox}). This image is fed into a deep segmentation network that classifies every pixel into one of several semantic terrain categories: woodland, waterway, road, building, and undifferentiated background. For the segmentation model, we used a pretrained DeepLabv3 \cite{chen2017rethinking} and fine-tune it on the LandCover AI dataset\cite{Boguszewski_2021_CVPR}.   

Finally, the two models are merged. The CV classifications are cross-referenced with the USGS-derived attributes to enhance the overall accuracy and consistency of the final terrain model. During the merge, the USGS data is used as the sole provider of elevation data and small-scale hydrography features (e.g., streams), while the segmented CV model is merged with the land-cover features to place buildings, woodlands, and roads onto the terrain. This is particularly helpful in areas where USGS data is less precise due to lower resolution. Large scale hydrography features in the segmented CV model, such as lakes, are validated against the USGS data prior to inclusion in the terrain model. While the CV model is trained to recognize water, it currently has a tendency to return false positives due to shaded areas in mountainous regions.

\section{Operational Challenges in Using the TDS} 
\label{sec:challenges} 
\begin{figure}
    \centering
    \includegraphics[trim={0 0 0 4.5cm},clip,width=0.9\linewidth]{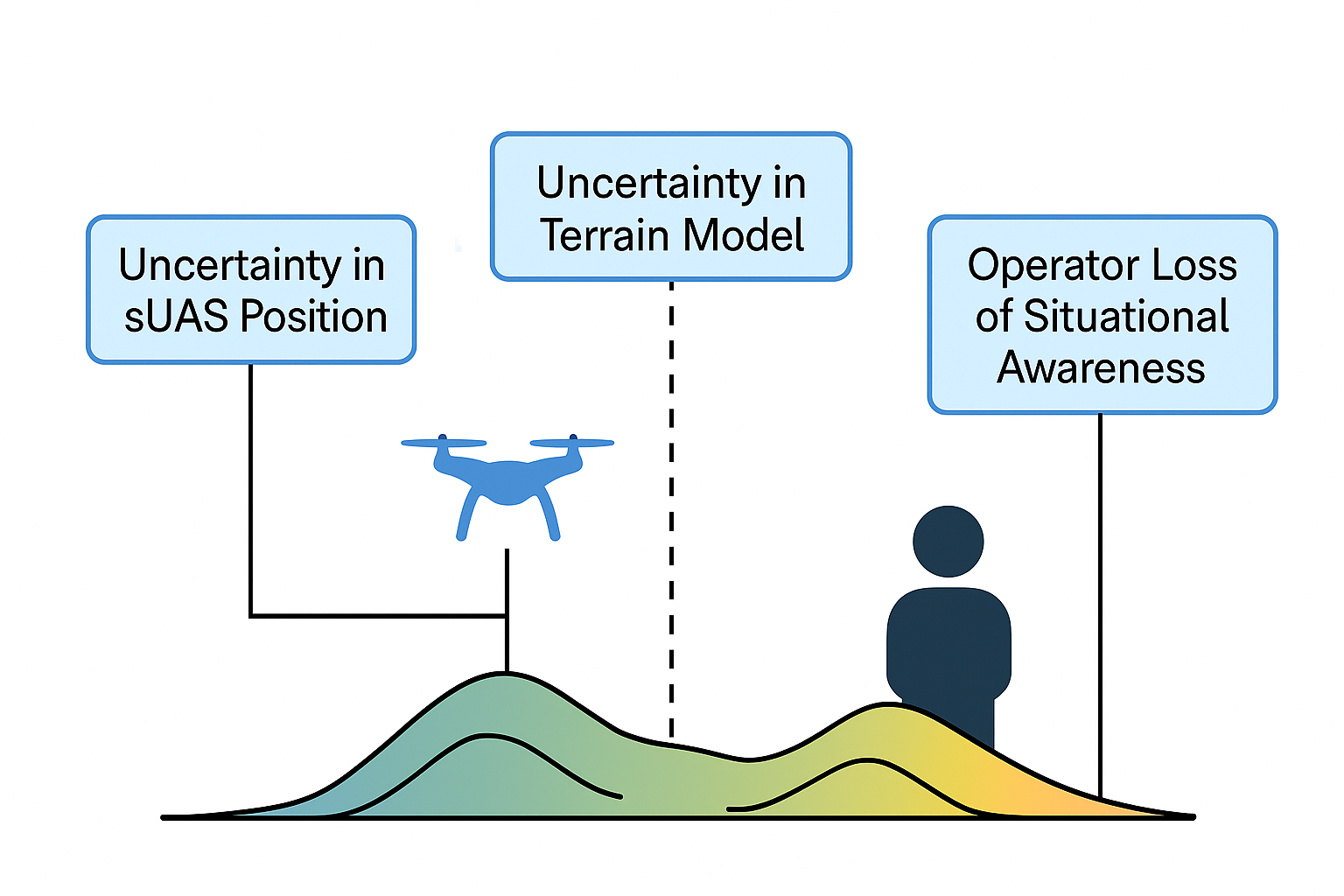}
    \caption{Illustration of terrain-based projection errors due to inaccuracies in model elevation, sUAS pose, and gimbal orientation. These errors compound to affect object localization, navigation, and situational awareness.}
    \label{fig:uncertainty-types}
    \vspace{-12pt}
\end{figure}

Terrain models provide a critical foundation for flight planning and safety in autonomous sUAS operations; however, their practical use introduces a range of challenges as summarized in Figure \ref{fig:uncertainty-types}. These stem not only from limitations in the models themselves, such as granularity, resolution gaps, or missing features, but also from the dynamic and uncertain nature of real-world sUAS deployments. In this section, we examine key operational challenges caused by these uncertainties.

\subsection{Terrain Model Challenges}
The first challenge concerns the accuracy of elevation and topographic detail within the terrain model. Misalignment with real-world features can result from low resolution, outdated datasets, or the failure to capture sharp structures such as ridges and crevices. This introduces three primary challenges:

\begin{itemize}
\setlength{\itemsep}{0.4em}
    \item [C1] {\bf Unexpected Obstacles:~} If the model omits elevated terrain features that occur between the collected data points, then sUAS may fly dangerously close to unknown obstacles, increasing the risk of collision.
    \item [C2] {\bf Inconsistent distance to the Ground:~} If the model smooths over depressions such as canyons or gullies, the actual distance to the terrain may be far greater than expected, reducing the effectiveness of onboard systems such as computer vision, which often depend on operating within a fixed range from the surface \cite{DBLP:conf/wacv/BernalSC24}.
    \item [C3] {\bf Inaccurate Object Geolocation:~} To geolocate an object in the physical world, we first extract the direction of the object from the sUAS, and then draw a ray towards it that eventually passes through a surface in the TDS. However, any inaccuracy in the geodesic planes will impact the accuracy of the geolocation, creating problems such as communicating coordinates of an object to other sUAS and to emergency responders on the ground.
\end{itemize}

These issues highlight that even a high-fidelity terrain model may introduce risk if it lacks the resolution or accuracy required by downstream systems.

\subsection{sUAS Challenges}
The TDS model provides critical support for the sUAS' perception, planning, and geolocation with accuracy impacted by the fidelity of the sUAS platform’s own sensing and pose estimation. Inaccuracies in position, orientation, or gimbal attitude can degrade alignment between the physical environment and the terrain model, and must therefore be carefully considered during testing. These platform-specific limitations introduce the following challenges:

\begin{itemize}
\setlength{\itemsep}{0.4em}
\item [C4] {\bf GPS Inaccuracies:~} sUAS often rely on GPS for absolute positioning, but GPS signals can be degraded by atmospheric conditions, multipath reflections in mountainous terrain, or temporary signal loss. Even modest errors of 1–3 meters can shift the projected location of an observed object significantly, especially when combined with uncertainty of the sUAS' roll, pitch, and yaw angles. 

\item [C5] {\bf Gimbal Sensor Accuracy:~} Camera gimbals are typically stabilized despite being physically mounted to the sUAS body and constrained by their available degrees of freedom. However, the digital model depends on accurate knowledge of the camera’s orientation, which is typically derived from sensors within the gimbal itself. Some gimbals lack sensors that directly measure their absolute heading or their orientation relative to the local tangent frame. Often the only available measurements are from angular position sensors associated with each controllable axis. Regardless of the sensor configuration, errors can arise from calibration drift, mechanical wear, control signal latency, or limited sensor resolution, distorting the estimated direction of observations, causing the projected ray used for geolocation to diverge from the actual line of sight, introducing further errors in object localization.

\item [C6] {\bf Visual Detection Uncertainty:~} Errors may also arise within the perception process, where bounding boxes generated by vision models may be imprecise due to inaccuracies of camera parameters, image resolution, field-of-view incurabilities, occlusion, or shallow viewing angles. Additionally, downstream software, such as the algorithm that computes the viewing direction, may introduce additional computational errors. Algorithmic errors may complicate the detection of underlying TDS problems.

\item [C7] {\bf Onboard Resource and Timing Constraints:~}  
Effective use of terrain models requires not just compute efficiency, but precise spatiotemporal alignment between sensor data and vehicle state. To project camera observations onto terrain, the system must accurately associate each video frame with the sUAS position, orientation, and gimbal angle at the exact timestamp. This requires high-frequency data logging, as well as either tightly synchronized sensors or interpolation procedures to estimate the vehicle’s state at arbitrary timestamps. Further, this data processing requirement may exceed the capacity of some onboard systems. Moreover, when terrain models are used for obstacle avoidance or terrain navigation, timing delays may force overly conservative behavior, such as reduced velocities or increased separation distances from terrain objects. In testing, such delays complicate the attribution of failure, making it difficult to distinguish between flaws in the terrain model and computational delays.
\end{itemize}

All of these limitations must be noted when validating the use of a terrain model under realistic operational conditions.

\subsection{Human Situational Awareness Challenges}
Finally, the presence of multiple interdependent uncertainties creates significant challenges for operator situational awareness, affecting the operator’s ability to interpret sUAS behavior, assess risk, and make timely decisions \cite{agrawal2020next, HERNANDEZ2025Navigating}.

\begin{itemize}

\item[C8] {\bf Recognizing and Interpreting Uncertainty in Context:~}
A human remote pilot must clearly understand the extent to which they can trust sUAS to operate safely and correctly with respect to TDS-related actions such as terrain avoidance and path-planning. However, the combined impact of uncertainties (as discussed above), creates a complex operational landscape where humans need to assess whether system behavior remains within acceptable bounds and to detect early signs of degradation before failure occurs. 
\end{itemize}
The system must therefore support effective situational awareness about the interactions between the sUAS and the TDS.

\subsection{Compounding Failures }
Individually, the challenges introduced by terrain model inaccuracies, sUAS sensing and processing limitations, and gaps in operator situational awareness can each degrade system performance. However, these issues rarely occur in isolation. In practice, failures often emerge from the interaction of all three components, producing outcomes that no single subsystem could have predicted or mitigated alone.
This underscores the need to validate not just individual components, but the integrated operational capability of the TDS. What matters is not only the accuracy of the underlying model, but how it functions in coordination with real sUAS behavior, handles uncertainty in sensing and geolocation, and enables meaningful human oversight during live missions.

\section{A TDS Testing Framework}
\label{sec:framework}
\begin{figure}[t]
    \centering
    \includegraphics[width=.8\linewidth]{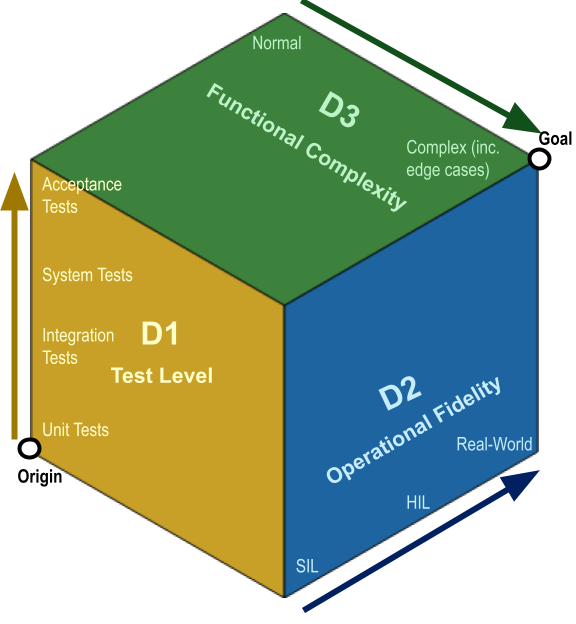}
    \caption{The TDS Testing Framework incorporates three dimensions focused on test levels (D1), operational fidelity (D2), and functional complexity (D3).}
    \label{fig:test-cube}
    \vspace{-8pt}
\end{figure}

To address these challenges, we present a testing framework to validate the Terrain-Aware Digital Shadow (TDS). The framework is designed to reveal weaknesses, guide system design and refinement, and assess whether the final system supports safe sUAS operation under real-world conditions. It is organized across three orthogonal dimensions of testing phase (D1), realism fidelity (D2), and functional complexity (D3) as depicted in Figure \ref{fig:test-cube}. 

\subsubsection*{D1: Test Levels}
The first dimension of the framework validates the system across increasing levels of test abstraction, adapting principles from the V-Model \cite{rook1986controlling}. This structured approach supports incremental verification from isolated software modules to full mission execution, and helps ensure that assumptions made during design are explicitly tested in realistic deployment conditions.

\begin{enumerate}
\setlength{\itemsep}{0.1em}
    \item \textit{Unit tests} assess and validate individual software components responsible for terrain parsing, altitude estimation, and coordinate projection. These tests are conducted in isolation using synthetic data and known inputs to verify algorithmic correctness and boundary behavior.

    \item \textit{Integration tests} evaluate whether terrain-informed modules—such as path planning, obstacle avoidance, or visibility estimation—operate correctly when interacting with other subsystems, such as the gimbal attitude. These tests expose interface mismatches, data formatting inconsistencies, or unintended coupling between terrain data and decision logic.

    \item \textit{System tests} examine how the TDS functions as part of a complete autonomy stack. This includes verifying that geolocation, trajectory planning, and environmental awareness function coherently in a mission context when terrain models are present or challenged by uncertainty.

    \item \textit{Acceptance tests} assess whether the TDS supports successful mission execution under real deployment conditions. These tests validate that model-informed behavior meets end-user expectations for safety, reliability, and situational awareness when operating in realistic environments and time constraints.

\end{enumerate}

\subsubsection*{D2: Simulation to Real-World Fidelity} 
The framework evaluates the TDS across increasing levels of fidelity, from software-in-the-loop (SIL), through hardware-in-the-loop (HIL), to real-world deployment. Each level contributes to identifying different classes of issues:

\begin{enumerate}
\setlength{\itemsep}{0.1em}
    \item \textit{SIL testing} enables fast, repeatable evaluation of core functions such as terrain parsing, elevation interpolation, coordinate transformation, and sensor alignment. These tests run on the generated terrain model in a noise-free environment, supporting idealized analysis.

    \item \textit{HIL testing} integrates real hardware components such as autopilots, GPS units, or gimbal sensors into a simulation-driven environment. Outputs of simulated components pass through physical interfaces to reveal integration issues such as latency, calibration drift, or model mismatches, all within a safe simulated context.

    \item \textit{Real-World testing} subjects physical sUAS to actual terrain features and environmental conditions. These tests expose edge cases that are difficult to simulate, identify failure modes that emerge only in physical-world conditions, and serve as the final stage of validating whether the TDS supports reliable and safe real-world operations.
\end{enumerate}

\subsubsection*{D3: Functional and Environmental Complexity} 
The third dimension captures two forms of complexity that evolve over the course of testing. The first reflects functional complexity, which increases through progressively more demanding scenarios, such as transitioning from simple waypoint following to dynamic path planning and inter-agent coordination. For example, one sUAS may geolocate a detected object and transmit its coordinates to another sUAS tasked with performing a follow-on action, requiring precise coordination and reliance upon shared data. The second reflects environmental complexity, progressing from well-mapped, relatively flat, open terrain to more difficult edge cases, including occluded ridgelines, dense vegetation, steep slopes, and GPS-denied zones. By evolving both aspects in tandem, the framework evaluates whether the TDS remains effective not only in controlled conditions but also under challenging and failure-prone scenarios.

\subsection{Selecting Effective Test Instances}
\label{SelectingEfectiveTestInstances}
As exhaustive testing that exercises every possible combination is not viable when considering cost and effort relative to return on investment, testers must select appropriate tests at each level. For example, during early stages of development, the developer is likely to conduct extensive unit and integration tests in simulation. As the project proceeds, engineers attempt to validate that features working in simulation also function correctly in HIL environments and, eventually, in the real world. A well-designed set of real-world integration tests, with validated intermediate values and successful outcomes, can often provide effective confirmation that outcomes from simulation or HIL testing are reproduced under operational conditions. Conversely, failed or inconsistent real-world integration tests may signal the need for additional lower-level testing, including real-world unit tests.

\subsection{Requirements-Driven Test Scenarios}
\begin{figure}[t]

    \centering
   \includegraphics[width=\columnwidth]{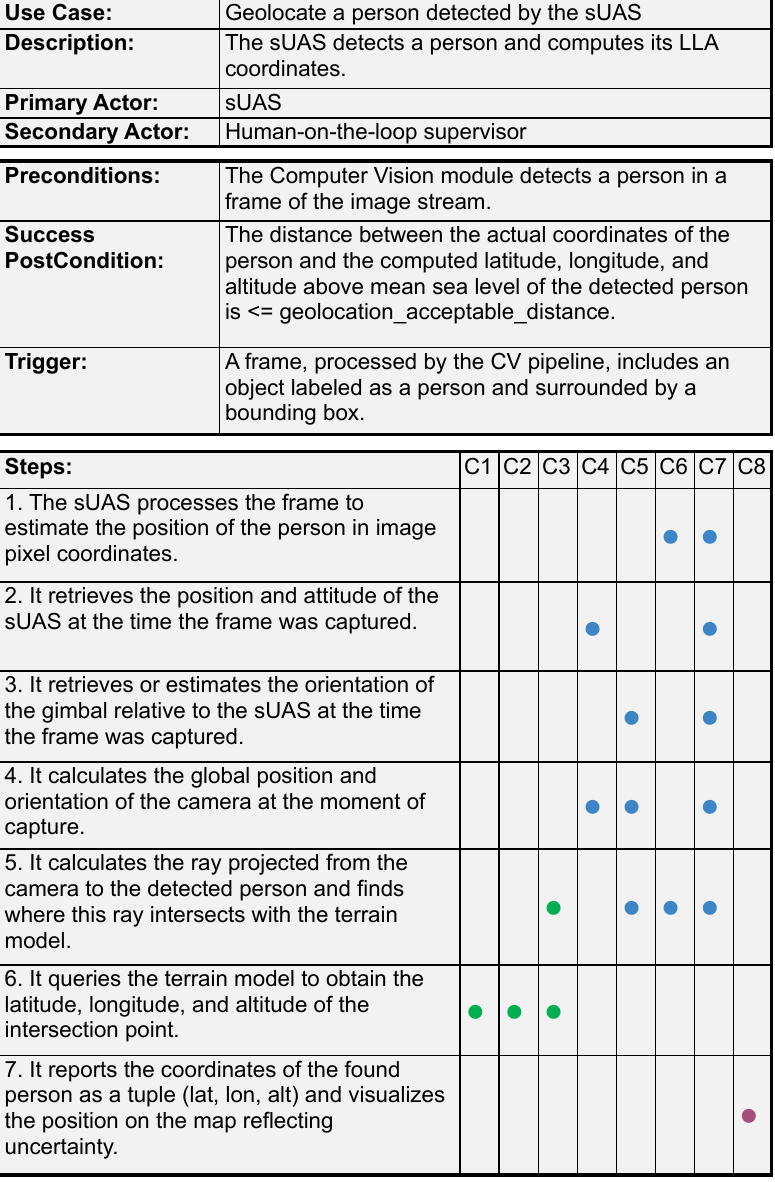}
    \caption{Use Case Scenario: Geolocating a person by casting a ray into the TDS.} 
    \label{fig:uc1}
\vspace{-12pt}
\end{figure}

To focus evaluation on critical system behavior, we adopt a scenario-driven testing approach in which sUAS interact meaningfully with the TDS. Rather than simply observing whether a test concludes successfully, each scenario step is explicitly mapped to one or more of the previously identified system challenges (e.g., C1–C8). This mapping helps ensure that test outcomes are interpreted in light of the specific uncertainties or limitations each scenario is designed to probe. Examples scenarios include the following:

\begin{itemize}
\item An sUAS uses onboard CV to geolocate an object, and then transmits the object's coordinates to another sUAS.
\item An sUAS flies to a waypoint over hilly terrain, while maintaining a stable altitude above the terrain for effective search and surveillance.
\item An sUAS scans the terrain for a place to land that is flat and devoid of trees, rocks, buildings, and water.
\end{itemize}
In the remainder of the paper we focus on the first example, as depicted in Fig. \ref{fig:uc1}, to evaluate specific challenges such as object geolocation accuracy (C3), GPS and pose uncertainty (C4, C5), and synchronization or resource constraints (C7).

\section{Applying the TDS Testing Framework}
\label{sec:applied}
The TDS testing framework was therefore progressively applied along each of the three dimensions for our targeted use-case (See Fig. \ref{fig:uc1}). In this section, we first describe the sUAS ecosystem, the testing environments along the different stages of operational fidelity (D2), and the different physical locations used to explore functional complexity (D3). We then present the tests and their results. Notably, several tests did not pass in their first cycle, requiring analysis of results, careful redesign of underlying algorithms, and subsequent re-testing.

\subsection{sUAS Ecosystem}
\begin{figure}[t]
    \centering
    \includegraphics[width=.9\linewidth]{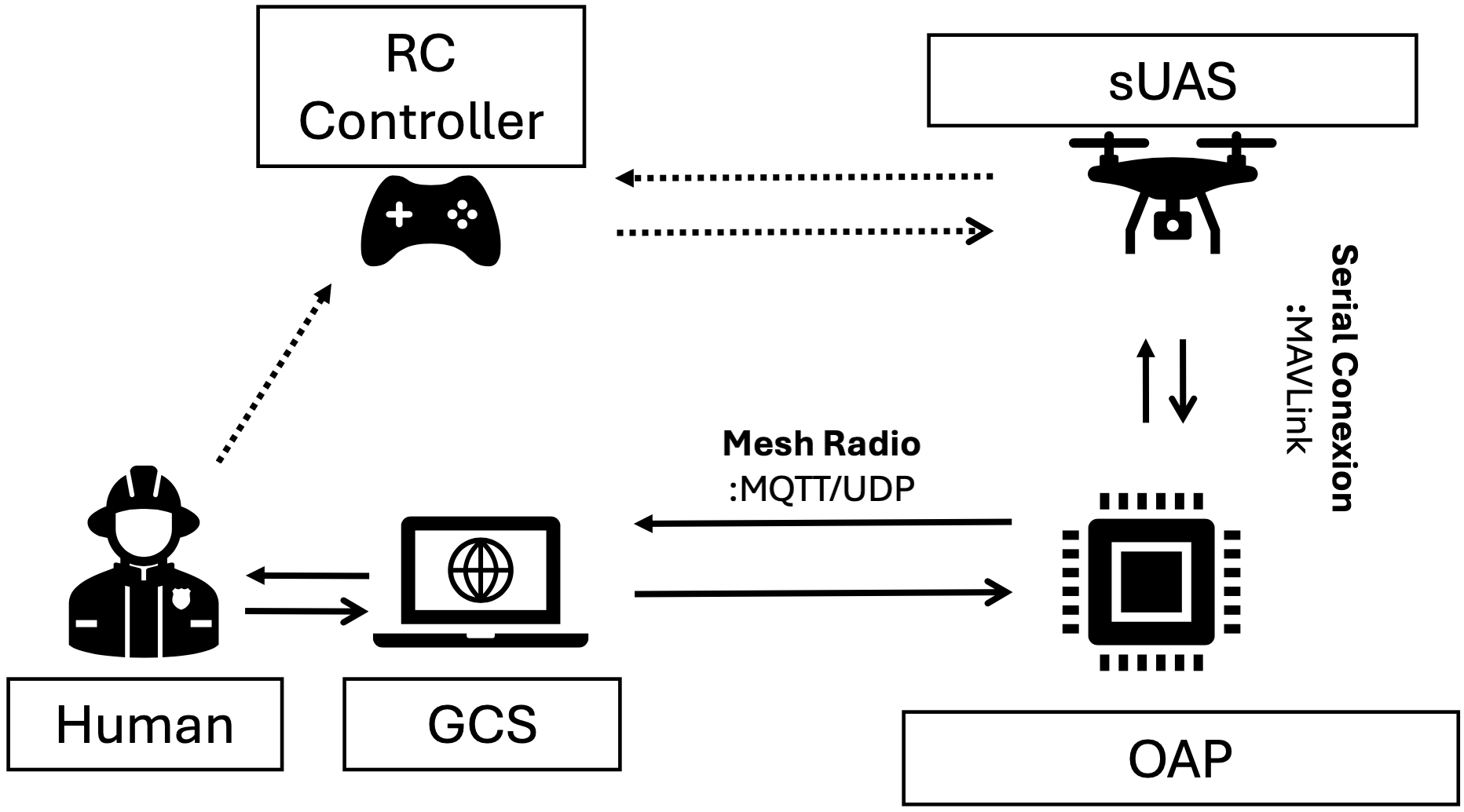}
    \caption{The sUAS ecosystem, showing the main components and the communications between them. The RC is used by the Human only under emergency scenarios.}
    \label{fig:ecosystem}
\end{figure}

\begin{figure}[t]
    \centering
    \includegraphics[width=.85\linewidth]{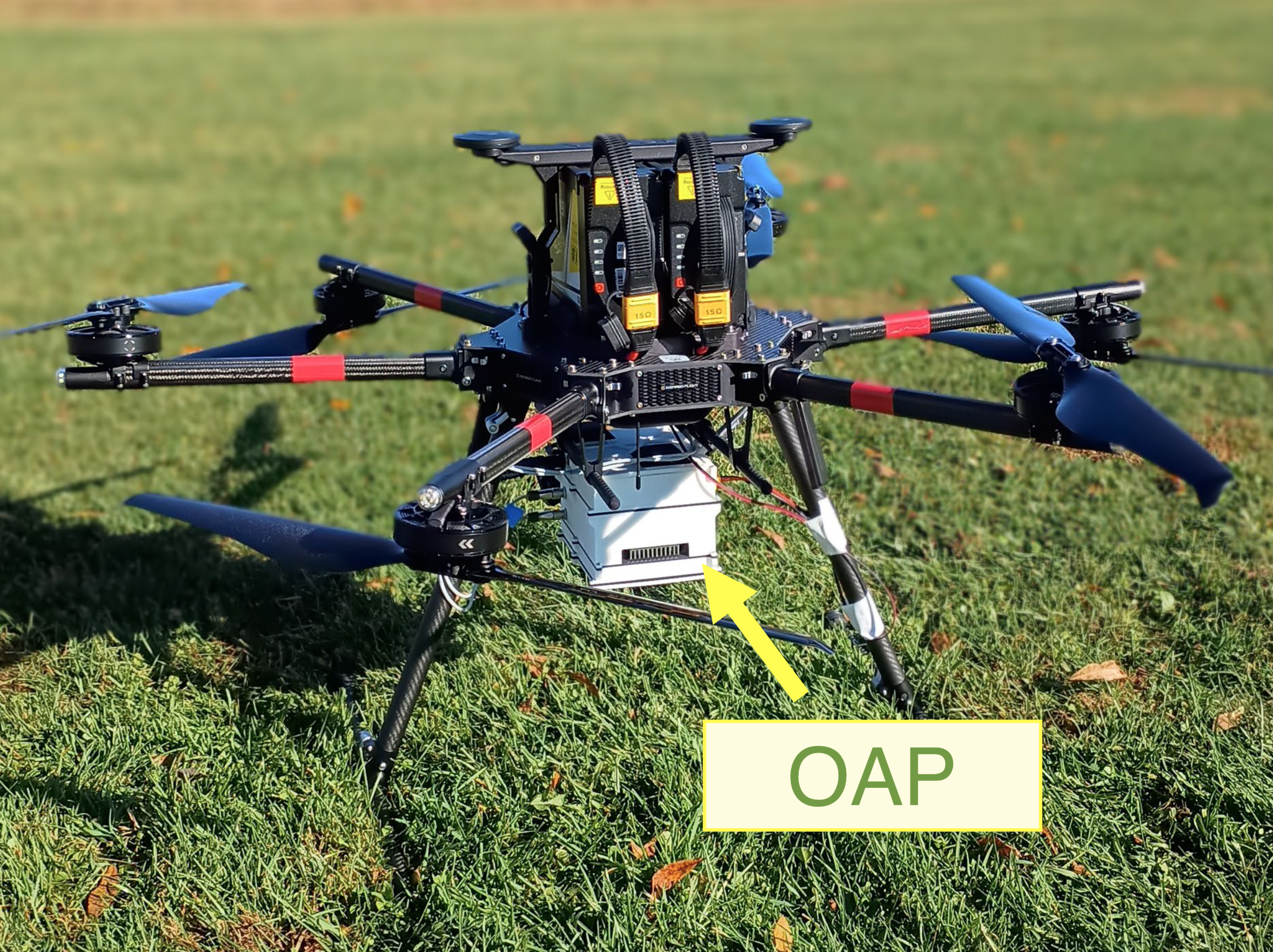}
    \caption{OAP attached to an Inspired Flight IF1200A. }
    \label{fig:sUAS}
    \vspace{-12pt}
\end{figure}

DroneResponse \cite{cleland2024human, chambers2024self, cleland2025cognitive, DBLP:conf/re/GranadenoBIC24}, our multi-sUAS ecosystem is comprised of four principal components: the sUAS (Drone), the Onboard Autonomous Pilot (OAP), the Ground Control Station (GCS) and the Human (see Fig. \ref{fig:ecosystem}). Every sUAS is equipped with our custom-built OAP, which communicates using the MAVLink protocol through a serial connection (see Fig. \ref{fig:sUAS}). The sUAS models used during these tests were two Inspired Flight IF1200A equipped with Ardupilot Flight Controller. The OAP uses a configurable ROS state machine which communicates with the core flight-controller to pilot the sUAS. It also hosts several critical components necessary for achieving geolocation. These include the onboard TDS, Camera, Gimbal and the CV models. The Gimbal is a Gremsy Two-Axis MIO capable of rotating in the pitch and roll direction; the yaw rotation is achieved by rotating the sUAS on the Z-axis. Communication between the OAP and the GCS is achieved using an MQTT and UDP Protocol through a mesh radio network. The GCS hosts additional services, such as our custom-made GUI and an Airleaser, responsible for authorizing airspace so that sUAS can avoid collisions in a preemptive manner. Humans supervise flights, maintaining situational awareness, and assuming  manual control of the sUAS using an RC Controller if needed. 

\subsection{Test Environments}
To achieve the required degrees of operational fidelity (D2)  we establish the following testing environments.

\subsubsection{Software-in-the-loop (SIL)} sUAS are simulated using Gazebo \cite{gazebo}, a simulation environment with high-fidelity physics. The OAP and GCS components are all containerized, enabling their deployment independent of specific hardware.  The simulated sUAS attitude and location can be streamed to QGroundControl (QGC), an open-source GCS software system to provide situational awareness during tests. 

\subsubsection{Hardware-in-the-loop (HIL)} 
\begin{figure}[t]
    \centering
    \includegraphics[width=.9\linewidth]{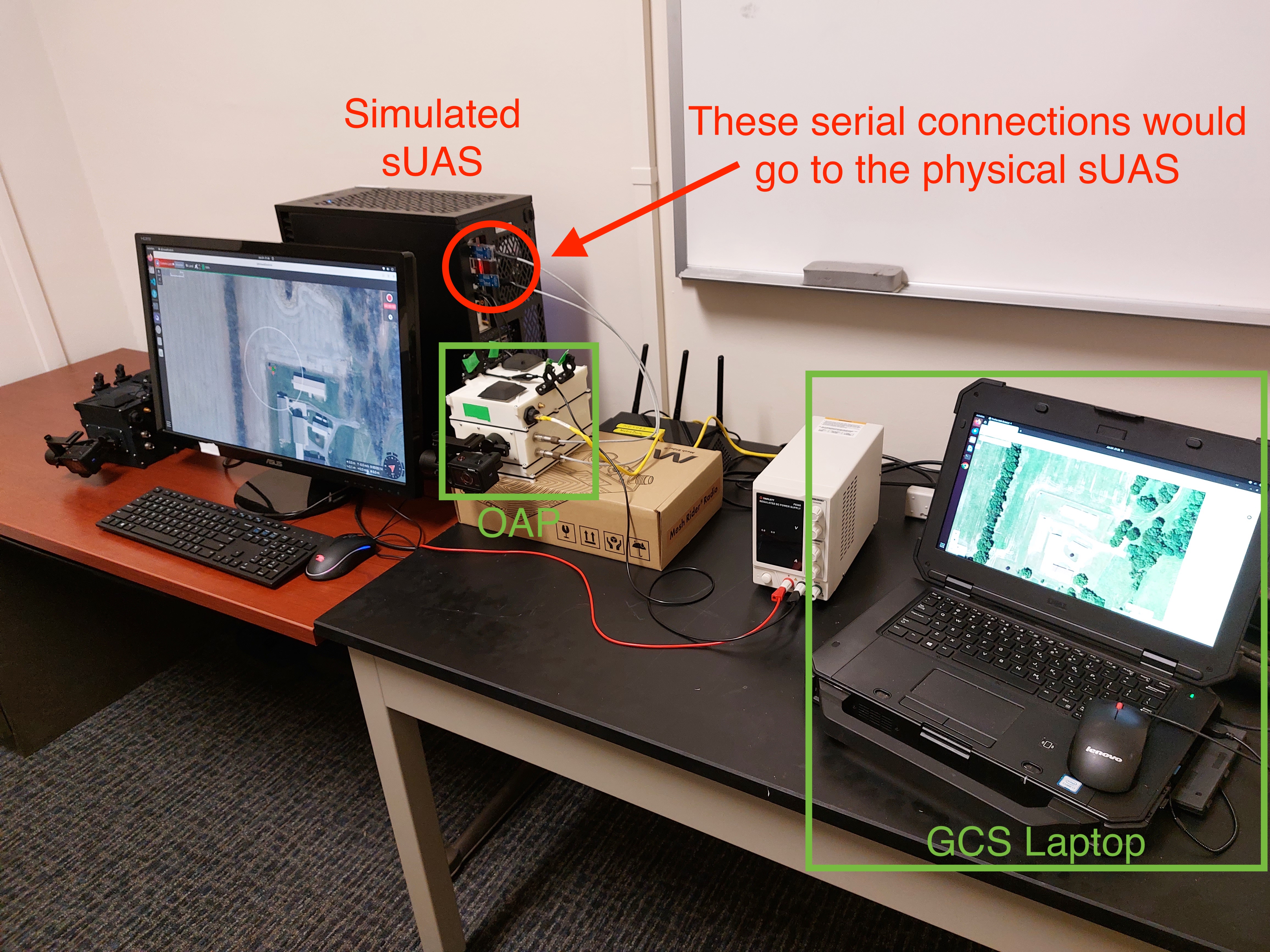}
    \caption{HIL setup, where the sUAS component is simulated, but the OAP and GCS components are the actual hardware.}
    \label{fig:lab_setup}
    \vspace{-12pt}
\end{figure}

sUAS are again simulated using Gazebo, but the OAP as well as the GCS use the same physical hardware components deployed in the field and shown in Fig. \ref{fig:lab_setup}. This is achieved by connecting the serial cables from an OAP to a PC running the simulation. Our HIL environment is supported by containerized OAP and GCS components, making it feasible to merge many SIL and HIL tests.

\subsubsection{Real-World}
Locations were chosen explicitly to test the functional complexity dimension (D3). Our primary real-world flight testing site was a farm in Southern Michigan, including (1) a relatively flat area with approximately 1m of terrain elevation delta (FARM-FLAT), and (2) a small gully that had at least 6m of elevation delta (FARM-GULLY). We also ran field tests at St. Patrick's Park in Northern Indiana, chosen for a relatively steep incline used as a toboggan hill (TOBOGGAN-HILL - $\sim$16m of elevation difference); however, we lacked authorization to fly in the park, and so these tests were all ground-based.  Additional locations during SIL and HIL tests included a diverse set of terrain types ranging from extreme mountainous conditions (Rockies), Hilly areas (Oklahoma), to several local sites in Indiana and Michigan.

\subsection{Executed Test Examples}
Tests progressed concurrently across the three dimensions; starting with SIL-based unit tests of basic features, and progressing to functionally complex, real-world, acceptance tests. We describe example tests here for illustrative purposes.\newline\vspace{-8pt}

\noindent{\textbf{Unit Tests (D1):~}}
Unit tests targeted individual components of the OAP, such as the TDS, Gimbal, and CV and were designed to address specific questions shown below with annotations explaining D2 and D3 levels:

\begin{table}[b]
    \centering
    \caption{Sample unit test performed by the TDS to validate that the model correctly retrieves LLA coordinates.}
    \begin{tabular}{|l|c|r|}
        \hline
        \textbf{Drone LLA}: (36.212189, -96.006905, 195, \\ DATUMREFERENCE.ELLIPSOIDWGS84) \\
        \hline
        \textbf{Camera Parameters:}: \\
            "fov\_h": 74,\\
            "image\_res": (1920, 1080), \\ 
            "target\_coords": (960, 810), \\
        \hline
        \textbf{Gimbal Quaternion}: (0.056115267, -0.0154703723, \\0.9608545, 0.27086953) \\
        \hline
        \textbf{Target LLA}: (36.21290054231726, -96.0083389306795, \\180.99648120246397)\\
        \hline
        \textbf{Retrieved LLA}: (36.21290054231724, -96.00833893067946, \\180.9964812024639)\\
        \hline
        \textbf{Description}: "Camera looks slightly down and to the north west." \\
        \hline
    \end{tabular}
    \label{tab:unit_test_tds}
\end{table}

\begin{itemize}[leftmargin=*]
    \item TDS [D2: SIL; D3: Varied]: Does the Terrain-Aware Digital Shadow (TDS) return correct terrain information given simulated gimbal orientations? Tests used input quaternions representing known gimbal orientations, and verified outputs from the TDS model. An example is shown in Table~\ref{tab:unit_test_tds}.
\item{Gimbal [D2: HIL; D3: Not applicable]}: Does the gimbal move to the specified pitch and roll angles, and do we correctly read back those angles from the hardware? Although a simulated gimbal is available in the Gazebo environment, the test target was the physical hardware, so tests were conducted in a Hardware-in-the-Loop (HIL) environment. A range of pitch and roll commands were issued to the gimbal, and the component was observed to follow them accurately, with reported angles matching the inputs.
\item{CV [D2: SIL; D3: Varied]}: Does the CV model detect persons from aerial views? CV models have previously been evaluated with SAR datasets \cite{DBLP:conf/wacv/BernalSC24, bernal2025psych, bernal2023hierarchically, maruvsic2018region, bovzic2019deep, sambolek2021automatic}. Both out-of-the-box YOLO11 model \cite{yolo11_ultralytics} and a custom YOLO11 model trained with NOMAD \cite{DBLP:conf/wacv/BernalSC24} were used during our real-world tests.
\end{itemize}

\noindent{\textbf{D1: Integration Tests (D1):~}} \label{integration}
Integration tests rely on the proper operation of the TDS with other components. For the scope of the current use case, we focus on the integration of the TDS, Gimbal and CV components of the OAP; however, the integration tests also use other previously validated GCS and OAP components such as the State Machine. 

\begin{figure}[t!]
    \centering
    \includegraphics[width=0.85\linewidth]{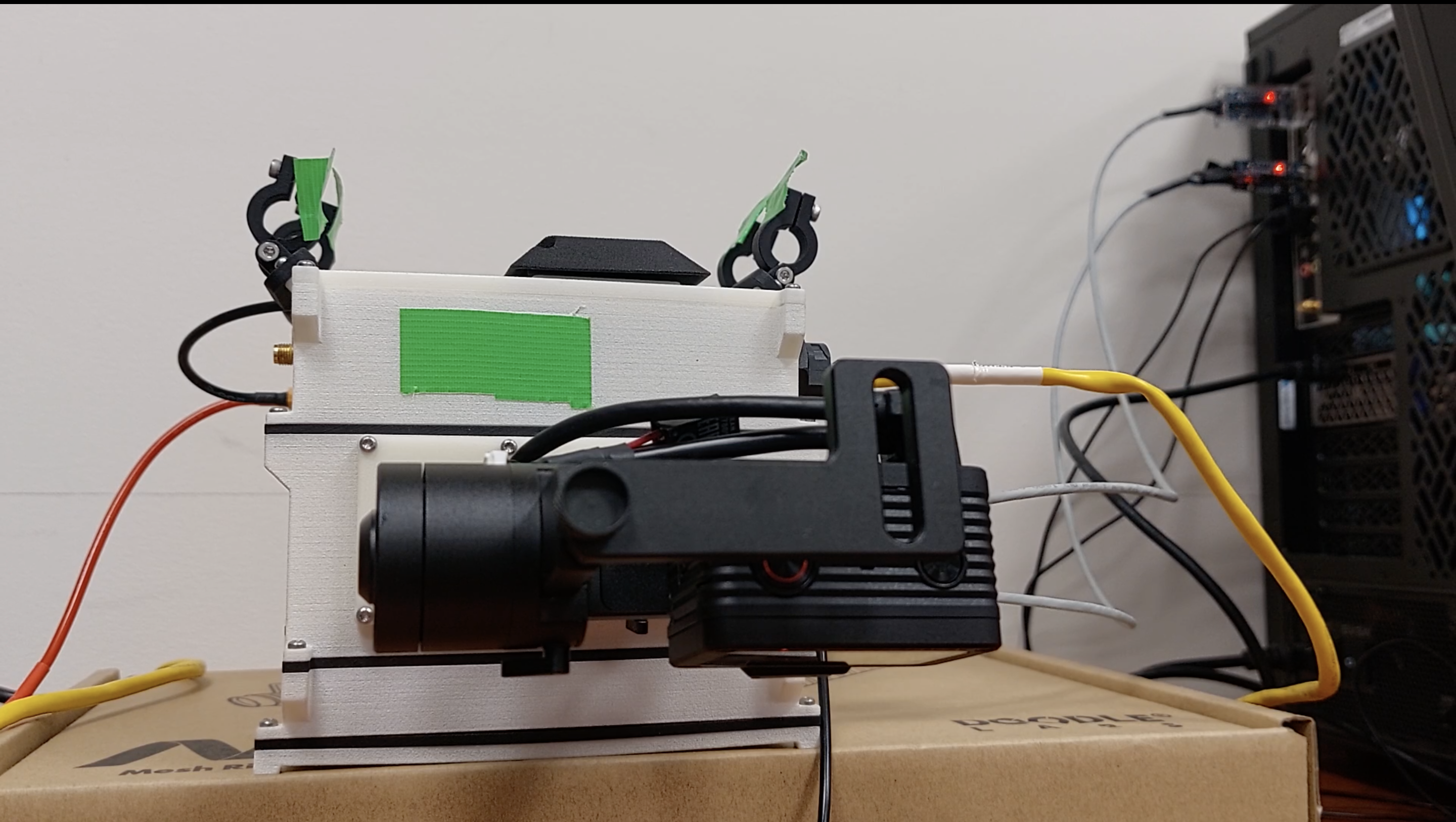}
        \caption{Gimbal position observed during HIL tests when looking almost directly above the target location.}
        \label{fig:subfig1}
    \label{fig:gimbal_hil}
    \vspace{-12pt}
\end{figure}

\begin{itemize}[leftmargin=*]
\item TDS with Gimbal [D2: HIL\&RealWorld, D3: Varied].
The geolocation accuracy of the TDS was tested for different gimbal positions, distances, and perspectives from a target LLA. The sUAS was first assigned a fixed stare point (LLA). The sUAS then queried the TDS for the LLA coordinates of the center pixel of its camera frame.  Experiments were first performed in a HIL environment and then in the real-world (D2).  For HIL tests we validated that the Gimbal moved appropriately when given a stare point (see Fig. \ref{fig:gimbal_hil}). Real-world tests were executed at both FARM locations (i.e., for relatively flat versus hilly terrain). 

\item TDS, Gimbal and CV [D2: HIL\&RealWorld, D3: Varied].
Geolocation of a detected person was tested as follows. The CV component detected a person, annotated it with a bounding box, and then geolocated the center pixel of the bounding box. Tests were conducted successfully in HIL and repeated at the FARM-FLAT, FARM-GULLY and TOBOGGAN-HILL. Results for both integration tests are  discussed in Section \ref{sec:discussion}.
\end{itemize}

\noindent{\textbf{System Tests (D1):~}}
Mission success in the multi-sUAS environment is only achieved through collaboration of multiple types and multiple instances of components (e.g., sUAS, OAP, GCS, Human) across the ecosystem.  

\begin{itemize}[leftmargin=*]
\item{Collaborative detection. [D2: SIL/HIL\&Real-World, D3: Simple]} In this test, one sUAS detects a person whilst flying to  waypoint with an initial gimbal pitch of 45 degrees. It publishes the coordinates of the detected person over MQTT, and a  second sUAS receives the coordinates. Both sUAS turn to face the detected person.  The test was executed in both HIL and at the FARM-FLAT site. A difference in latency from the OAP's MQTT message broker was noticed between HIL and FARM-FLAT deployments; while the test repeatedly passed the HIL environment, the difference in latency  caused some message loss when deployed in the real-world, requiring modifications to the code base receiving the messages. This represents a clear example of the HIL to Real-Wold shift.  
\end{itemize}

\noindent{\textbf{Acceptance Tests (D1)~}} To date, we have not conducted user acceptance tests of the TDS environment, as additional functional tests are needed first. However, during the testing process we identified several areas in which we (as users) lacked situational awareness about current uncertainty levels and their impact on task outcomes and sUAS safety. Designing for situational awareness, and ultimately validating this aspect of the TDS is out of scope for this paper, but is an essential part of our longer-term validation plan.

\subsection{Results and Discussion} \label{sec:discussion}
The application of the TDS Testing Framework provided interesting insights into the performance of the TDS for geolocation purposes.
The SIL and HIL validation stages enabled system debugging prior to the significant cost and effort required in field deployments. These costs include logistics of identifying test locations, travel, and both hardware and software system setup and tear-down in the field. In particular, the HIL tests allowed us to validate critical components such as the Gimbal, in advance of real-world flights.

\begin{figure}[t]
    \centering
    \begin{subfigure}[t]{0.48\columnwidth}
        \centering
        \includegraphics[trim={0 0 0 5cm},clip,width=\linewidth]{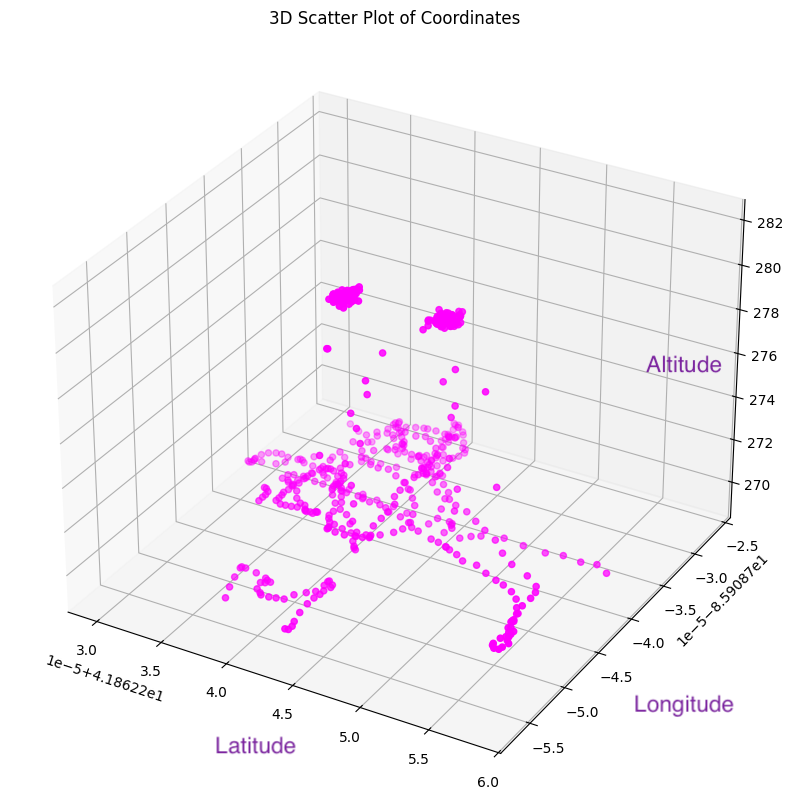}
        \caption{Fuchsia sUAS hovering twice.}
        \label{fig:test_c_location_fuchsia}
    \end{subfigure}
    \hfill
    \begin{subfigure}[t]{0.48\columnwidth}
        \centering
        \includegraphics[trim={0 0 0 5cm},clip,width=\linewidth]{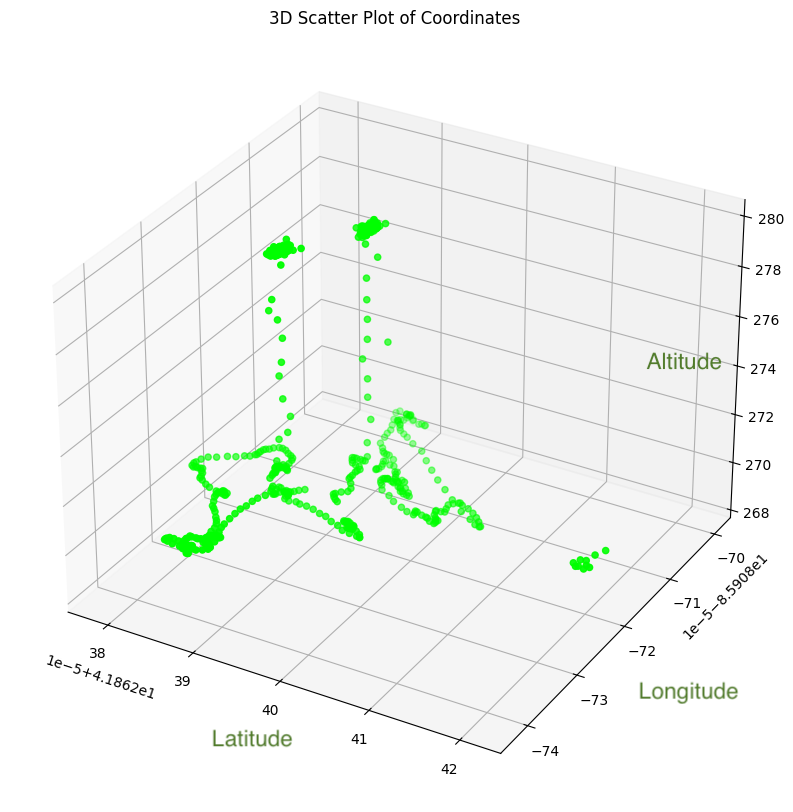}
        \caption{Lime sUAS hovering twice.}
        \label{fig:test_c_location_lime}
    \end{subfigure}
    \caption{LLA coordinates of two different sUAS during a test where they were instructed to hover twice at the same spot. Both sUAS show high GPS uncertainty on the ground, which improves once airborne.}
    \label{fig:test_c_location}
\end{figure}

\begin{figure}[t!]
    \centering
    \begin{subfigure}[t]{0.48\columnwidth}
        \centering
        \includegraphics[trim={0 0 0 0.8cm},clip,width=\linewidth]{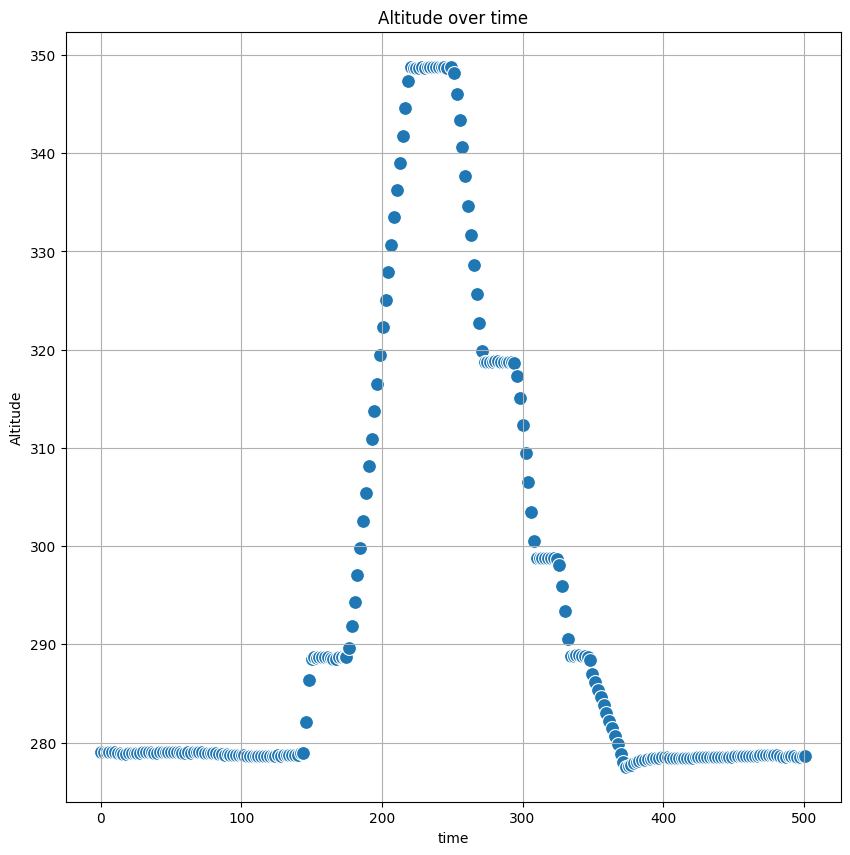}
        \caption{Fuchsia flying from a believed ground altitude of 280m [amsl].}
        \label{fig:test_altitude_fuchsia_b}
    \end{subfigure}
    \hfill
    \begin{subfigure}[t]{0.48\columnwidth}
        \centering
        \includegraphics[trim={0 0 0 0.8cm},clip,width=\linewidth]{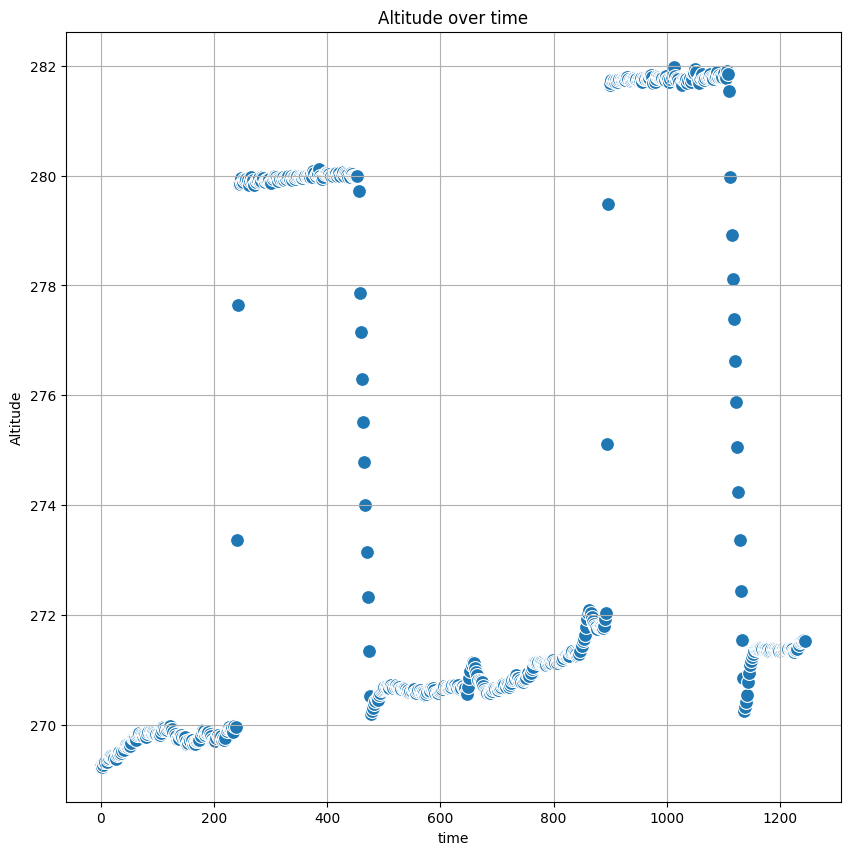}
        \caption{Fuchsia hovering twice from a believed ground altitude of 270m and 272m [amsl].}
        \label{fig:test_altitude_fuchsia_c}
    \end{subfigure}
    \caption{Altitude over time of the same sUAS during two different tests on the same day, showing a significant jump in reported altitude while on the ground, relative to the TDS baseline of 274–275m at this location.}
    \label{fig:test_altitude_fuchsia}
\end{figure}

We now discuss observations and results from the field-tests, all of which were characterized by GPS uncertainty. Fig \ref{fig:test_c_location} illustrates a simple test in which two different sUAS were each tasked with hovering twice. The figure shows high uncertainty of the sUAS location  on the ground ($\sim$4m for `Fuchsia', $\sim$2.5m for `Lime'\footnote{All of our drones use colors (e.g., `Lime' or `Fuschia' as unique IDs)}), with considerable decrease in uncertainty once in the air ($\sim$1m for Fuchsia, $\sim$0.3m for Lime). The graph also shows that both sUAS landed in slightly different locations from their launch locations.  
In contrast, altitude uncertainty was observed both on the ground and in the air during all tests. Fig \ref{fig:test_altitude_fuchsia} shows altitude readings from the same sUAS during two different tests on the same day and same location (FARM-FLAT) where elevation readings were expected to be  274-275m above mean seal level (amsl) according to the TDS. Both tests showed persistent altitude deviations from the TDS and from each other.

\begin{figure}[htbp!]
    \centering
    \begin{subfigure}{0.4\textwidth}
        \centering
        \includegraphics[trim={0 0 0 0},clip,width=\linewidth]{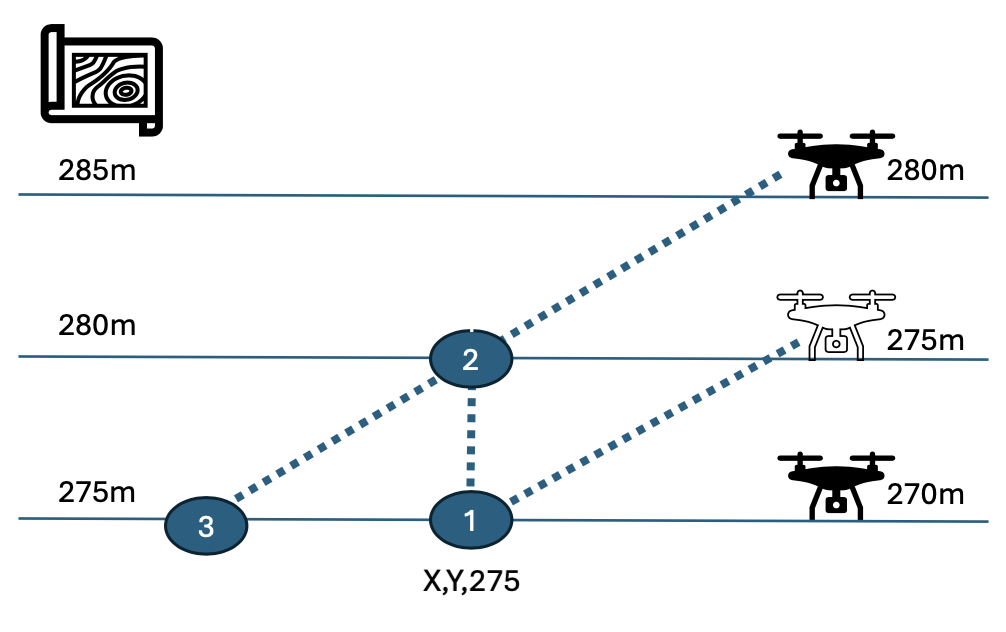}
        \vspace{-0.5cm}
        \caption{Stare point error produced by the altitude error of the sUAS. The sUAS will be staring at point (3) when it was expected to stare at point (1).}
        \label{fig:geolocation_mismatch_1}
        \vspace{0.2cm}
    \end{subfigure}
    
    \begin{subfigure}{0.4\textwidth}
        \centering
        \includegraphics[trim={0 0 0 0},clip,width=\linewidth]{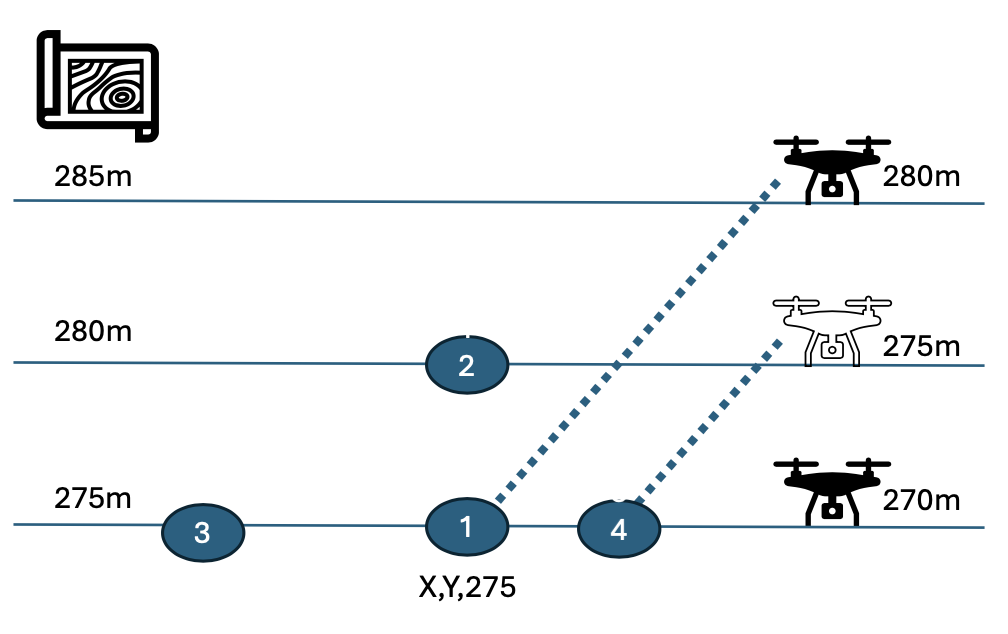} 
        \vspace{-0.5cm}
        \caption{Geolocation error produced by the altitude error of the sUAS. When a person is detected at point (1), the sUAS creates a ray from its position towards point (1), but from the TDS perspective the sUAS will be casting the ray from a lower altitude, returning point (4) as the coordinates of the person.}
        \label{fig:geolocation_mismatch_2}
    \end{subfigure}
    \caption{Errors produced by the difference in elevation between the TDS and the sUAS. The right side of the image describes sUAS' beliefs while the left side describes TDS' beliefs.}
    \label{fig:geolocation_mismatch}
    \vspace{-12pt}
\end{figure}

Throughout the first set of integration tests (TDS with Gimbal), we observed that geolocation coordinates were offset by a few meters in different directions. To quantify this, we used the Haversine formula \cite{van2017heavenly} and found that the average latitude–longitude error across all experiments was approximately 1.5m, with a maximum of around 4.2m.  In contrast, elevation error remained below 1m, even in FARM-GULLY tests where elevation ranged from 268m to 274m above mean sea level. Both results are strong given the various sources of uncertainty described in Section~\ref{sec:challenges}. 
Fig \ref{fig:geolocation_mismatch_1} analyzes the altitude measurements displayed in Fig. \ref{fig:test_altitude_fuchsia_c} and shows how the mismatch in sUAS altitude readings affects the stare point location. When given the latitude, longitude and altitude (LLA) of a stare point (1), from the sUAS perspective, that point corresponds to point (2), which is actually in the air; therefore, the center of the frame from the camera perspective becomes point (3).  Nonetheless, when projecting a ray from the center of the camera frame we still retrieve point (1), indicating that the TDS retrieves the correct LLA given a ray projection from the camera, but the obtained LLA coordinates do not exactly match the expected coordinates given a point in an image. This is explored in the following tests.

\begin{figure}[htbp!]
    \centering
    \begin{subfigure}{0.4\textwidth}
        \centering
        \includegraphics[trim={0 0 0 0},clip,width=\linewidth]{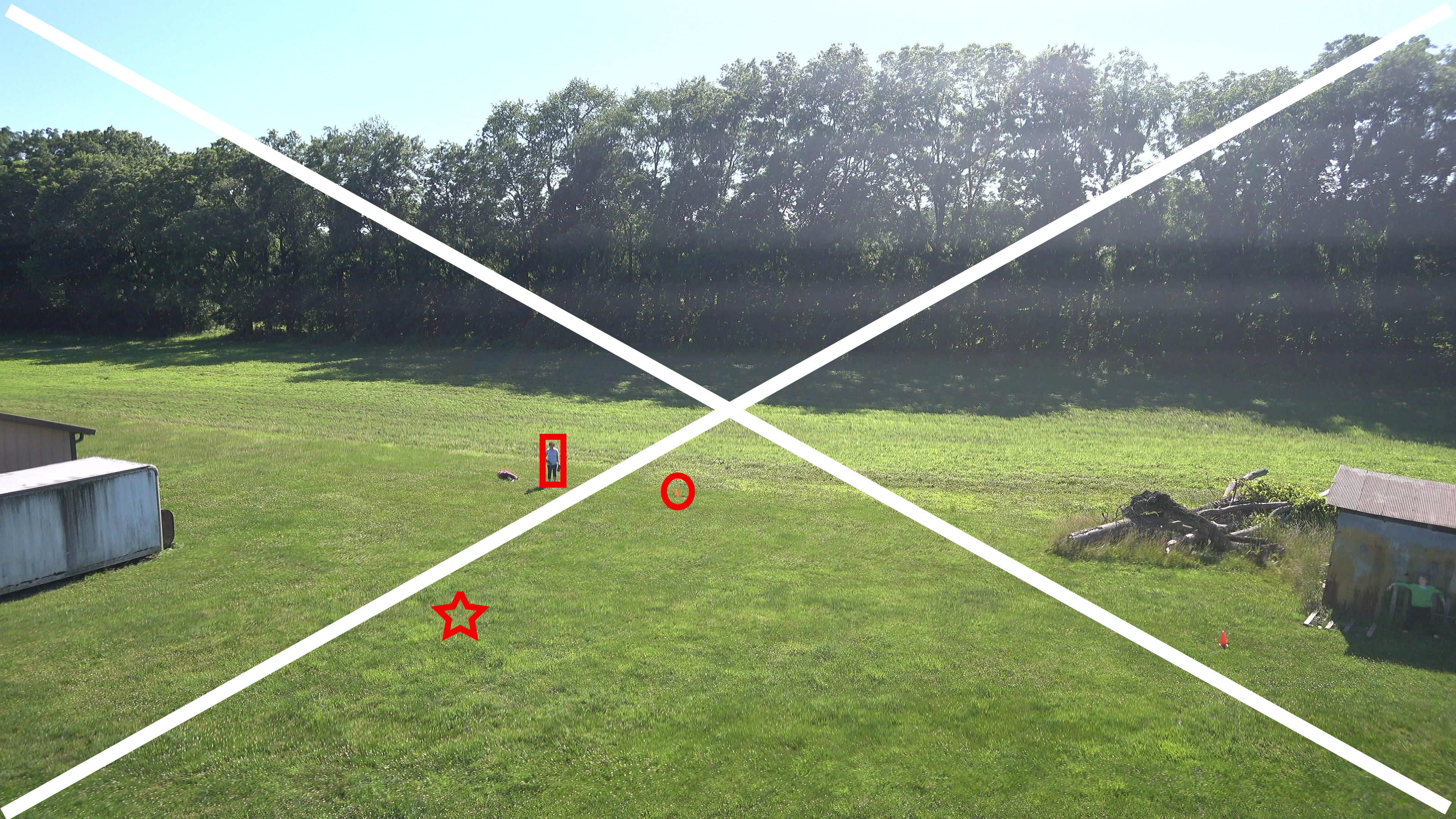}
        \vspace{-0.5cm}
        \caption{}
        \label{fig:sample_image_1}
        \vspace{0.2cm}
    \end{subfigure}
    
    \begin{subfigure}{0.4\textwidth}
        \centering
        \includegraphics[trim={0 0 0 0},clip,width=\linewidth]{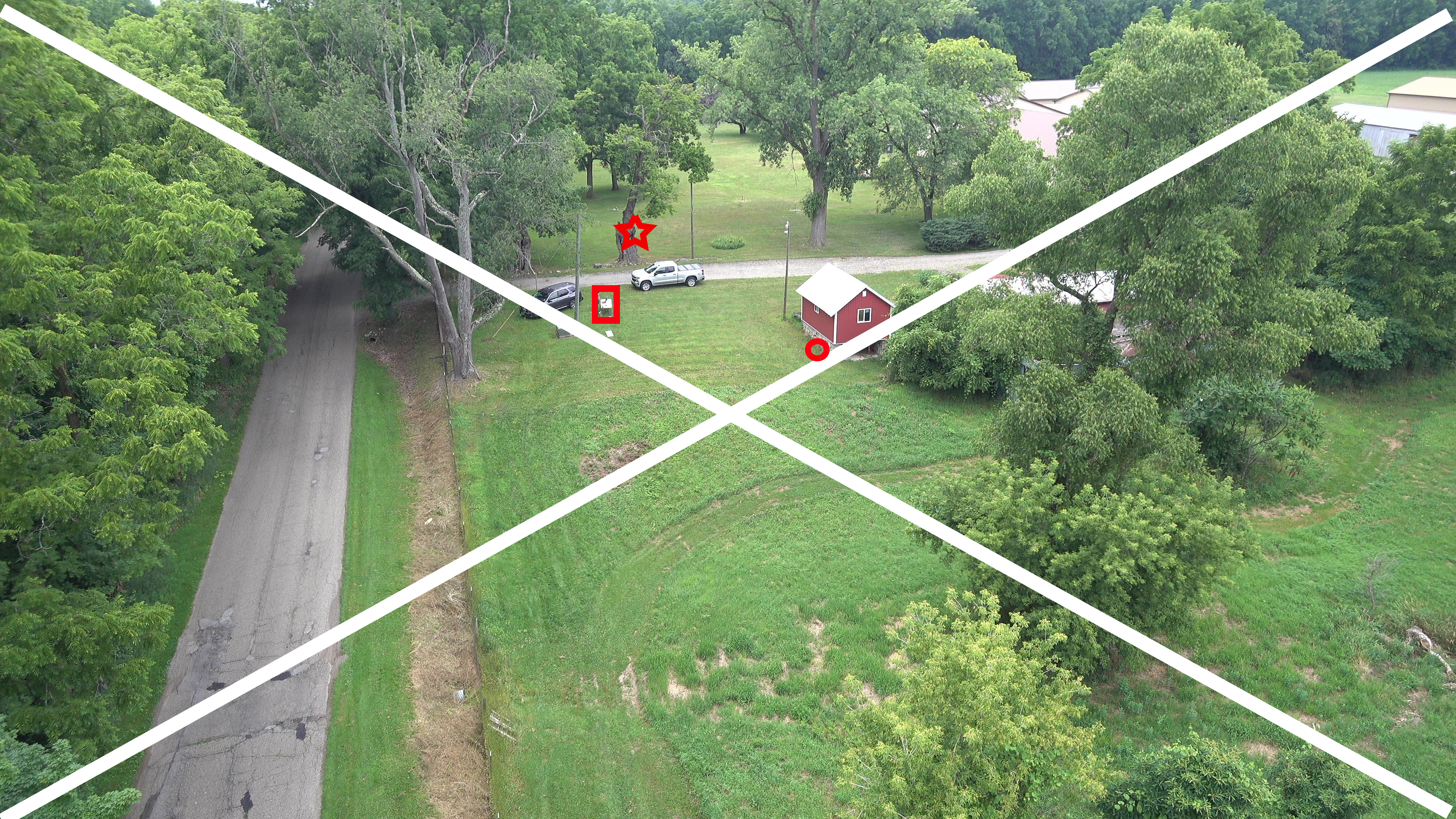} 
        \vspace{-0.5cm}
        \caption{}
        \label{fig:sample_image_2}
    \end{subfigure}
    \caption{Annotated frames show the specified (red circle) versus the actual stare point (white cross), and the detected person (red rectangle) vs the geolocation coordinates returned by the TDS model (red star). These results show how mismatched altitude readings between the sUAS and TDS affect the stare point and impact geolocation accuracy.}
    \label{fig:sample_image}
    \vspace{-12pt}
\end{figure}

Our second set of tests, retrieving geolocation coordinates of a detected person based on the geolocation of the center of a bounding box, introduced additional uncertainties. We observed that sUAS' altitude error causes a geolocation error for a detected person in an image.  Fig \ref{fig:geolocation_mismatch_2} exemplifies a situation where the sUAS reads a lower altitude measurement compared to the TDS.  In this scenario, the sUAS stares at point (3) when it is expected to stare at point (1). Furthermore, when a person is detected at point (1), the sUAS creates a ray from its current position towards point (1), but from the TDS perspective the sUAS casts the ray from a lower altitude, returning point (4) as the coordinates of the person.  This behavior was confirmed when reviewing the recordings and analyzing the frames obtained from the camera. For example Fig \ref{fig:sample_image_1} shows this particular scenario, where the sUAS believes itself to be at a lower altitude with respect to the TDS during a test performed at FARM-FLAT.  The circle represents point (1), while the cross represents point (3).  The person was detected near point (1) but the geolocation response from the TDS was closer to the sUAS (marked with a star).  The inverse situation happens when the sUAS believes itself to be at a higher elevation with respect to the TDS, shown in Fig \ref{fig:sample_image_2} during a test performed at FARM-GULLY.  Additionally, these figures show the effect of latitude and longitude uncertainty. For example, in Fig \ref{fig:sample_image_2} the sUAS should have yawed more towards the right, however, during one set of tests at FARM-GULLY where high HDOP values had been observed (indicating satellite geometry problems), the sUAS consistently flew further to the left than expected (almost crashing into the trees at one point), indicating that it believed its position to be further to the right than it actually was. When validating geolocations during these  tests, we observed geolocation errors of over 10m. These, and other similar errors highlight the need for further unit tests through  cyclic testing, where observations made in integration tests at the field lead to modifications to the code-base supported by additional unit tests in SIL or HIL environments.

These insights are invaluable in helping us to understand weaknesses of the TDS, identify where uncertainties impact performance, and ultimately to design solutions to improve the outcome of TDS-dependent use-cases. 

\section{Threats to Validity}
\label{sec:threats}
There are several threats to validity. First, our experimental work has focused on only one use case while other scenarios such as flying over hilly terrain have not yet been evaluated. Nonetheless, the use-cases share many underlying characteristics related to correct geolocation of the sUAS in relation to the TDS. Second, although the framework defines three dimensions of test level, fidelity, and scenario complexity, these may not explicitly capture all factors relevant to TDS validation. We treat additional aspects as variants embedded within the functional complexity dimension. Third, while the challenge categories (C1–C8) provide a structured basis for interpreting test outcomes, they do not constitute a complete taxonomy. New types of operational challenges may emerge in extended missions, under atypical conditions, or with larger swarms. As such, both the dimensions and challenge categories should be regarded as adaptable and subject to refinement as new scenarios and platforms are introduced.
In a fourth limitation, all field experiments were conducted using two Inspired Flight 1200A drones in a geographical area with hills but no mountains. These limitations constrain our generalizability claims, and future tests will involve a wider range of platforms and more diverse terrain.

Despite these limitations, the application of the framework across SIL, HIL, and real-world testing facilitated systematic planning and interpretation. The process successfully uncovered system-level issues and supported their mitigation, demonstrating the framework’s practical value.
\section{Related Work}
\label{sec:related}

Testing frameworks for autonomous systems have advanced across dimensions such as simulation fidelity, scope, and environmental complexity. A key trend is the shift away from exclusive reliance on field testing, which is often costly and impractical, and toward high-fidelity simulation and digital twin environments \cite{heiko2024}. These simulations reduce time and cost by using physics-based engines and virtual replicas, but their value depends on how accurately they reflect reality. Digital twins, which stay synchronized with real-world data, help meet this need. For example, Matalonga et al. used a digital twin of a sUAS vision system to identify hazards and worst-case collisions in simulation, with results validated in flight \cite{matalonga2024}. Hybrid setups are also emerging. Wang et al. developed a testbed combining real and simulated sUAS in a shared space with close temporal and spatial alignment \cite{wang2023design}, helping bridge the gap between simulation and reality.

Another key dimension in the validation landscape is testing scope, which ranges from unit-level verification to full-system acceptance testing. Verifying autonomous systems requires a multi-tiered approach, where components are first validated in isolation and then integrated for system-level testing \cite{heiko2024}. Many frameworks follow a staged hierarchy similar to traditional V-models. In the sUAS domain, this typically moves from Software-in-the-Loop (SIL) to Hardware-in-the-Loop (HIL) and then to flight testing \cite{jiang2025stepbystepguidecreatingrobust}. SIL enables early functional checks, HIL introduces real hardware effects, and flight trials expose system behavior in the field. This layered strategy catches issues early and manages complexity \cite{jiang2025stepbystepguidecreatingrobust}. Pikner et al. emphasize the need for both component-level and system-level V\&V due to the blend of deterministic and stochastic elements in autonomy \cite{heiko2024}. While widely practiced, these stages are rarely unified under a single framework, such as the 3-dimensional framework proposed in this paper.

The third dimension addresses terrain and environmental complexity, highlighting the importance of scenario-based and terrain-aware testing. As autonomous sUAS operate in more diverse environments, validation must consider topographic variation, weather, and rare edge cases. Since exhaustive real-world testing is infeasible \cite{mustafasa2021}, researchers have turned to smarter scenario generation methods. Karunakaran et al. propose a data-driven approach that models real-world driving scenarios to produce realistic edge cases for autonomy testing \cite{s24010108}. In the sUAS domain, simulations increasingly integrate real geographic data. Valencia et al. developed an open-source digital twin of an Andean mountain region in ROS/Gazebo, enabling rehearsal in high-altitude terrain \cite{valencia2025open}. Their tests showed strong alignment with real flight trajectories, even under wind gusts. These efforts underscore the value of incorporating environmental complexity into validation workflows to expose system weaknesses under high-risk conditions.

In summary, prior work has advanced simulation fidelity, layered testing strategies, and scenario-based validation \cite{heiko2024, matalonga2024, jiang2025stepbystepguidecreatingrobust, mustafasa2021, s24010108, valencia2025open}. However, most frameworks treat these dimensions separately. The proposed 3-Dimensions framework unifies test scope, fidelity, and terrain complexity into a single model. Evaluating sUAS systems across all three dimensions offers more comprehensive validation than approaches addressing each dimension in isolation.

\section{Conclusions}
\label{sec:conclusions}
This paper has presented a structured testing framework for evaluating terrain-aware digital shadows in autonomous sUAS systems. By organizing tests across testing levels, operational fidelity, and functional complexity, and by explicitly linking test steps to operational challenges, the framework enables systematic and meaningful validation of TDS-supported behavior within a digital twin context.

Simulation plays a critical role during early development, but many challenges such as terrain occlusion, GPS drift, and sensor misalignment are difficult to observe or reproduce in simulation alone. By combining simulation, hardware-in-the-loop, and real-world testing, the framework supports more thorough and realistic evaluation, identifying areas in which real-world uncertainties impact performance, and helping to ensure that systems are not only well-designed but also reliable in practice.

Field testing revealed several areas for improvement. These included the need for tools that support rapid test generation in dynamic conditions, integration of Built-In Test (BIT) mechanisms to detect errors early, and real-time diagnostic outputs to assist with in-field troubleshooting. Additionally, ensuring that Remote Pilots in Charge maintain situational awareness, especially in degraded environments, is essential for safe and effective test execution. These insights will guide future refinements of both the framework and supporting infrastructure.

\balance
\bibliographystyle{IEEEtran}
\bibliography{digitwin}

\end{document}